\documentclass[showkeys,showpacs,nofootinbib,superscriptaddress]{revtex4-2}
\usepackage{amsmath}
\usepackage{amssymb}
\usepackage{graphicx}
\usepackage{float}
\usepackage{xcolor}
\usepackage{physics}
\usepackage{ulem}

\allowdisplaybreaks

\begin{document}

\title{
Fermion condensate at the event horizon
}

\author{
Vladimir Dzhunushaliev
}
\email{v.dzhunushaliev@gmail.com}
\affiliation{
Department of Theoretical and Nuclear Physics,  Farabi University, Almaty 050040, Kazakhstan
}

\affiliation{
Institute for Experimental and Theoretical Physics, Farabi University, Almaty 050040, Kazakhstan
}

\affiliation{
Academician J.~Jeenbaev Institute of Physics of the NAS of the Kyrgyz Republic, 265 a, Chui Street, Bishkek 720071, Kyrgyzstan
}

\author{Vladimir Folomeev}
\email{vfolomeev@mail.ru}
\affiliation{
Academician J.~Jeenbaev Institute of Physics of the NAS of the Kyrgyz Republic, 265 a, Chui Street, Bishkek 720071, Kyrgyzstan
}


\begin{abstract}
Some arguments are considered in favor of the idea that the canonical anticommutation relations for fermions should be modified in curved spacetime near the event horizon of a black hole. Such a modification is expected to lead to a change in the source term of the inhomogeneous Dirac equation describing the two-point Green’s function. By introducing an {\it ad hoc} source into the Dirac equation that mimics the modification of these anticommutation relations, stationary solutions are obtained and interpreted as two-point Green’s functions of fermions located near the event horizon. Owing to their stationarity, these Green’s functions describe a fermion condensate near the event horizon.
\end{abstract}


\date{\today}

\maketitle

\section{Introduction}

The event horizon is one of the most intriguing predictions of general relativity. It possesses remarkable properties that lead to various physical consequences.
In particular, the event horizon is a null hypersurface separating the region from which no future-directed causal signal can reach future null infinity.
 Another property is that in quantum field theory (QFT), the presence of an event horizon gives rise to the hypothetical Hawking radiation \cite{Hawking:1974rv}; this signifies that QFT in curved spacetime near the event horizon differs significantly from QFT in flat spacetime. The foundation of any quantum theory rests on the assertion that physical quantities must be represented by operators acting on a certain vector space, whose elements are quantum states describing a given quantum physical system. If the Hawking radiation does indeed exist, it implies that the properties of field operators near the event horizon must differ substantially from their properties in flat spacetime. Consequently, the following question arises: do other novel physical phenomena exist near the event horizon?

Here, we demonstrate that a fermionic field condensate can arise near the event horizon. To this end, we consider the equation for the two-point Green's function of the fermionic field, where the source term on the right-hand side is a certain function that simulates the modification of the field operator properties in curved spacetime near the event horizon. This modification stems from the fact that anticommutation relations in curved spacetime must differ from the canonical anticommutation relations in flat spacetime. In Minkowski space, the standard anticommutation relations lead to the well-known expression for the source (in the form of the Dirac delta function) on the right-hand side of the Dirac equation describing the propagator (the two-point Green's function). In curved spacetime, the alteration of the operator properties modifies the anticommutation relations, thereby changing the right-hand side of the Dirac equation for the two-point Green's function. The aim of our study is to investigate the properties of such a Dirac equation with an {\it ad hoc} source on the right-hand side, which mimics the modification of anticommutation relations in curved spacetime near the event horizon.

A similar idea has already been considered in quantum mechanics, where the Heisenberg uncertainty principle is modified in one way or another: the so-called ``Generalized Uncertainty Principle.'' In Ref.~\cite{Maggiore:1993rv}, it was shown that the introduction of the Generalized Uncertainty Principle leads to a minimal length on the order of the Planck length naturally emerging from any quantum theory of gravity. In Ref.~\cite{Maggiore:1993kv}, the most general form of the
$\kappa$-deformed algebra of the coordinate operator $\hat x$ and momentum operator $\hat{p}$
is considered. Ref.~\cite{Capozziello:1999wx} discusses the Generalized Uncertainty Principle within the framework of quantum geometry.

The study of the properties of the Dirac equation in various black hole backgrounds has a long history, with the results of these investigations compiled in the book \cite{Chandrasekhar}. These studies are dedicated to the propagation of waves in different black hole spacetimes. Research in this direction remains active today. For instance, Ref.~\cite{Malik:2024nhy} investigates the quasinormal modes of massless Dirac field perturbations around black holes.

There are also intriguing results regarding self-consistent solutions to the Einstein-Dirac equations. Spherically symmetric solutions to these equations were likely first obtained in Ref.~\cite{Finster:1998ws}. This topic was further developed in Refs.~\cite{Herdeiro:2017fhv,Herdeiro:2019mbz}, where spherically and axially symmetric solutions in Einstein-Dirac gravity were found, as well as in Refs.~\cite{Konoplya:2021hsm,Dzhunushaliev:2025lki,Dzhunushaliev:2025ntr}, which presented a wormhole supported by gravitating spinor and electromagnetic fields.
Also, Ref.~\cite{Dzhunushaliev:2025dma} discusses the physical effect of condensation of a classical spinor field at the event horizon.

\section{Propagator in QFT}

The standard procedure for obtaining a propagator in QFT is as follows. One introduces the creation, $\hat{a}^\dagger_{\vec{p} \lambda}$,
and annihilation, $\hat{a}_{\vec{p} \lambda}$, operators for a fermion  $f^-$, and the creation, $\hat{b}^\dagger_{\vec{p} \lambda}$, and annihilation, $\hat{b}_{\vec{p} \lambda}$,
operators for an antifermion  $f^+$ satisfying the canonical anticommutation relations
\begin{align}
	\left\lbrace 
		\hat{a}^\dagger_{\vec{p} \lambda}, \hat{a}_{{\vec{p}}^\prime \lambda^\prime}
	\right\rbrace & = \delta_{\vec{p} \vec{p}^\prime} \delta_{\lambda \lambda^\prime}, 
\label{anti_1}\\
	\left\lbrace 
		\hat{b}^\dagger_{\vec{p} \lambda}, \hat{b}_{{\vec{p}}^\prime \lambda^\prime}
	\right\rbrace & = \delta_{\vec{p} \vec{p}^\prime} \delta_{\lambda \lambda^\prime}, 
\label{anti_2}\\
	\left\lbrace 
		\hat{a}_{\vec{p} \lambda}, \hat{a}_{{\vec{p}}^\prime \lambda^\prime}
	\right\rbrace & = 
	\left\lbrace 
		\hat{a}^\dagger_{\vec{p} \lambda}, \hat{a}^\dagger_{{\vec{p}}^\prime \lambda^\prime}
	\right\rbrace = 
	\left\lbrace 
		\hat{b}_{\vec{p} \lambda}, \hat{b}_{{\vec{p}}^\prime \lambda^\prime}
	\right\rbrace = 
	\left\lbrace 
		\hat{b}^\dagger_{\vec{p} \lambda}, \hat{b}^\dagger_{{\vec{p}}^\prime \lambda^\prime}
	\right\rbrace = 0 . 
\label{anti_3}
\end{align}
Subsequently, one can define the vacuum expectation value of the product of field operators $\hat{\psi}$, 
\begin{align}
	\imath S_+ \left( x_1 - x_2\right) & = 
	\left\langle 
		0 \left| 
		\hat{\psi}^\dagger \left( x_1 \right) \hat{\bar{\psi}}\left( x_2 \right) 
		\right| 0 
	\right\rangle = 
	\left( \imath \gamma^\mu \partial_\mu + m 
	\right) \int \frac{d \vec{p}}{\left( 2 \pi\right)^3 } 
	\frac{e^{- \imath p \left(  x_1 - x_2\right) }}{2 \epsilon_p} ,
\label{S+}
\end{align}
and similarly for $S_-$. 

The propagator $S_+$ satisfies the Dirac equation with a source
\begin{equation}
	\left( \imath \gamma^\mu \partial_\mu - m 
	\right) S_+ = \imath \delta\left( x_1 - x_2 \right) . 
\label{propagator_S+}
\end{equation}
The propagator  $S_{\pm}$ 
for fermions and the corresponding propagator for gauge fields are employed in perturbative QFT to investigate interacting fields, as is done, for instance, in the Standard Model of interacting fields. 
However, such a procedure fails for strongly interacting fields, such as those in quantum chromodynamics (QCD), where, for example, the confinement problem arises due precisely to the strong interaction between quarks. All this implies that, in QCD for instance, the two-point Green's function
\begin{equation}
	G\left( x_1, x_2\right) = 
	\left\langle 
		0 \left| 
		\hat{\psi}^+ \left( x_1 \right) \hat{\bar{\psi}}\left( x_2 \right) 
		\right| 0 
	\right\rangle
\label{green}
\end{equation}
is not the propagator defined by the expression \eqref{S+}, but rather must be determined by an infinite set of nonperturbative Dyson-Schwinger equations for all possible Green's functions of quarks $q$
and the gauge potential $A^a_\mu$. 
 This implies that the operator $\hat{\psi}$ is not determined by the anticommutation relations \eqref{anti_1}--\eqref{anti_3}.

This leads to the following conclusion: \textit{the operators of strongly interacting fields possess properties distinct from those of noninteracting field operators described by the anticommutation relations \eqref{anti_1}--\eqref{anti_3}.}

This allows us to conjecture that the same phenomenon occurs when fundamental (gauge and fermionic) fields interact with a strong gravitational field: \textit{the operators of these fields will differ from those in flat spacetime.}
That is, the fermionic field operators will satisfy anticommutation relations distinct from \eqref{anti_1}--\eqref{anti_3}. Here, it should be noted that a similar idea of modifying commutation relations (within the context of quantum mechanics) has already been considered in studies dedicated to the so-called ``Generalized Uncertainty Principle.''

Another scenario where field operators may differ from those in flat spacetime is the event horizon, near which a purely quantum phenomenon~-- the hypothetical Hawking radiation~-- potentially arises. This radiation possesses a strictly quantum nature and demonstrates substantial differences between QFT in gravity (near the event horizon) and the properties of QFT in flat spacetime. It can be assumed that the presence of an event horizon will also lead to other effects that are absent in flat spacetime. Here, we explore such a possibility by assuming that the canonical anticommutation relations for the fermionic field change significantly near the event horizon, leading to a modification of the equation for the two-point Green's function (referred to as the propagator in flat spacetime). As we demonstrate below, such a modification of the anticommutation relations can result in the emergence of a fermionic field condensate near the event horizon.

Another argument in favor of this viewpoint is the position of W. Heisenberg, which he articulated in Ref.~\cite{heis} (chapter 2, \S2): ``The initial or the final states are, by definition, states in which the interaction between the particles can be neglected. Such states, and only such states, can be constructed by applying sums of products of asymptotic operators on the vacuum. They are usually supposed to form a complete set of states in the sense that any process of interaction will finally, when the interaction has become negligible, lead to a state within this set. Even if this is true, the state of interaction itself may belong to a wider group of states which cannot be represented by the asymptotic states alone.'' 
This viewpoint of W. Heisenberg should be understood to mean that the properties of strongly interacting field operators fundamentally differ from those of weakly interacting fields; therefore, the description of strongly interacting fields must be inherently distinct from the methods developed for weakly interacting fields.

\section{Two-point Green's function $G\left( t, \vec{r}_H; t, \vec{r} \right) $ near the event horizon
}
Let us consider the two-point Green's function $G\left( t, \vec{r}_H; t, \vec{r} \right) $
near the event horizon. Our primary assumption is that a fermion condensate exists near the event horizon. Such a condensate constitutes a stationary configuration in the spherically symmetric gravitational field of a Schwarzschild black hole, depending on the time $t$
and the difference $(\vec{r} - \vec{r}_H)$ as follows:
\begin{equation}
	G_{\alpha \beta} \left( t, \vec{r}_H; t, \vec{r} \right) = G _{\alpha \beta}\left(t,  \vec{r} - \vec{r}_H \right) = 
	\left\langle Q \left| 
		\hat{\psi}^\dagger_\alpha \left( t, \vec{r}_H \right) \hat{\psi}_\beta \left( t, \vec{r} \right) 
	\right| Q \right\rangle ,
\label{Green_func}
\end{equation}
where $\left. \left| Q \right. \right\rangle $ is a certain quantum state describing a given physical system. Just as in the case of QFT in Minkowski space, the Green's function $G_{\alpha \beta} \left(t,  \vec{r} - \vec{r}_H \right) $
with two spinor indices $\alpha, \beta$ must satisfy the Dirac equation with a source term  $J$ on the right-hand side:
\begin{equation}
	\imath \left( \gamma^\mu\right)_{\beta \gamma} 
	\left( G_{\alpha \gamma}\right) _{;\mu} - \tilde{m}  G_{\alpha \beta} = J_{\alpha \beta} ,
\label{Dirac_Schwarz}
\end{equation}
where $\tilde m$ is the mass of the spinor field and the semicolon denotes the covariant derivative defined as
$
	G_{; \mu} = D_\mu G = 
 [\partial_{ \mu} +1/8\, \omega_{a b \mu}\left( \gamma^a  \gamma^b- \gamma^b  \gamma^a\right)] G 
$.
Here $\gamma^a$ are the Dirac matrices in the standard representation in flat space  [see, e.g.,  Ref.~\cite{Lawrie2002}, Eq.~(7.27)] and an {\it ad hoc} source $j$, 
which mimics the modification of canonical anticommutation relations, is a spinor. In turn, the Dirac matrices in curved space, $\gamma^\mu = e_a^{\phantom{a} \mu} \gamma^a$, are derived  using the tetrad
 $ e_a^{\phantom{a} \mu}$, and $\omega_{a b \mu}$ is the spin connection [for its definition, see Ref.~\cite{Lawrie2002}, Eq.~(7.135)].
 
 We seek the Green's function $G_{\alpha \beta} \left(t,  \vec{r} - \vec{r}_H \right) $ in the following form:
 \begin{align}
 \begin{split}
 	G_{\alpha \beta} \left(t,  \vec{r} - \vec{r}_H \right) & =  
 	\xi^\dagger_\alpha \chi_\beta , 
\\ 
 	\chi & = \frac{1}{2} e^{-\imath  \tilde{\Omega} t}
 	\begin{pmatrix}
 	 -\imath e^{\frac{1}{2} \imath (\theta -\varphi )} \tilde{u} \left( r - r_H\right) 	\\
 	 \imath e^{-\frac{1}{2} \imath (\theta +\varphi )} \tilde{u} \left( r - r_H\right) 	\\
 	 -e^{-\frac{1}{2} \imath (\theta +\varphi )} \tilde{v} \left( r - r_H\right) 						\\
 	 e^{\frac{1}{2} \imath (\theta -\varphi )} \tilde{v}\left( r - r_H\right) 								\\
 	\end{pmatrix} , 
 	 \end{split}
 \label{Green_anzatz}
 \end{align}
 where $\xi^\dagger_\alpha$ is a constant spinor. The source
  $J_{\alpha \beta} \left(t,  \vec{r} - \vec{r}_H \right)$ is taken in a similar form
\begin{align}
\begin{split}
	J_{\alpha \beta} \left(t,  \vec{r} - \vec{r}_H \right) & = 
  \xi^\dagger_\alpha j_\beta , 
\\ 
	j & =	- \frac{1}{2} e^{-\imath  \tilde{\Omega} t}
 	\begin{pmatrix}
 	 -\imath e^{\frac{1}{2} \imath (\theta -\varphi )} \tilde{j}_1 \left( r - r_H\right) 	\\
 	 \imath e^{-\frac{1}{2} \imath (\theta +\varphi )} \tilde{j}_1 \left( r - r_H\right)  \\
 	 -e^{-\frac{1}{2} \imath (\theta +\varphi )} \tilde{j}_2 \left( r - r_H\right) 	 					\\
 	 e^{\frac{1}{2} \imath (\theta -\varphi )} \tilde{j}_2 \left( r - r_H\right) 		
 	\end{pmatrix} .
\end{split}  
\label{source}
\end{align}
In Eq.~\eqref{Dirac_Schwarz}, the Dirac operator $\left(\imath \gamma^\mu D_\mu - \tilde{m}\right)$ 
acts on the spinor $\chi$, which allows us to cancel the constant spinor $\xi^\dagger$
in this equation. Here, it is necessary to emphasize a fundamental difference between the Green's function \eqref{Green_anzatz} in curved spacetime near the event horizon and the Green's function \eqref{S+} in flat Minkowski space. In Minkowski space, the Green's function vanishes outside the light cone, i.e., in spacelike directions. In our case, however, the Green's function \eqref{Green_anzatz} is assumed to describe a stationary 
fermion condensate and therefore takes nonzero values in these directions.

The Dirac equation \eqref{Dirac_Schwarz} for the two-point Green's function $G\left(t,  \vec{r} - \vec{r}_H \right) $
in a Schwarzschild black hole background with an {\it ad hoc}
source mimicking the modification of the anticommutation relations for the fermionic field operators $\hat{\psi}$
takes the following dimensionless form:
\begin{align}
	\sqrt{\frac{x}{x+1}} \, v' 
	+ \frac{1 + \left(1 + \frac{1}{4 x}\right) \sqrt{\frac{x}{x+1}}}{x+1} v 
	+\left(m - \frac{\Omega }{\sqrt{\frac{x}{x+1}}}\right) u 
	& = j_1 , 
\label{Dirac_dim_less_1}\\
	\sqrt{\frac{x}{x+1}} \, u' 
	+ \frac{- 1 + \left(1 + \frac{1}{4 x}\right) \sqrt{\frac{x}{x+1}}}{x+1} u 
	+\left(m + \frac{\Omega }{\sqrt{\frac{x}{x+1}}}\right) v 
	& = j_2 . 
\label{Dirac_dim_less_2}
\end{align}
Here, the prime denotes the derivative with respect to $x$, and the following dimensionless quantities are introduced:  $x = r/r_H - 1$, 
$
	v = r_H^{3/2} \tilde{v}, u = r_H^{3/2} \tilde{u}, m = r_H \tilde{m}, \Omega =  r_H \tilde{\Omega}, 
	j_{1,2} =   r_H^{5/2} \tilde{j}_{1, 2}
$. 

To obtain regular solutions to the Dirac equations \eqref{Dirac_dim_less_1} and \eqref{Dirac_dim_less_2} describing the two-point Green's function \eqref{Green_func}, it is necessary to select appropriate functions $j_{1,2}(x)$ that mimic the modification of the fermionic field operator properties near the event horizon, and to investigate the behavior of the functions $v(x)$ and $u(x)$ near the event horizon.

In the following subsections, we consider various cases for the values of the spinor functions $v(x)$ and $u(x)$ at the event horizon, which consequently require different types of the sources $j_{1,2}$.

\subsection{The case of nonzero $v(x)$ and $u(x)$ at the event horizon
}
\label{non_zero}

In this case, we take {\it ad hoc} sources $j_{1,2}(x)$ in the following form: 
\begin{align}
	j_1 & = e^{-\Sigma x} \left( 
		\frac{q_0}{\sqrt{x}}  + q_1 
	\right) , 
\label{source_anzatz_1}\\	
	j_2 & = e^{-\Sigma x} \left( 
		\frac{p_0}{\sqrt{x}}  + p_1 
	\right) , 
\label{source_anzatz_2}
\end{align}
where $\Sigma = \sqrt{m^2 -\Omega^2}$ and $q_{0,1}$ and $p_{0,1}$ are
some arbitrary parameters chosen in such a way as to eliminate the singular terms on the left-hand side of the Dirac equations \eqref{Dirac_dim_less_1} and \eqref{Dirac_dim_less_2}, which diverge at the event horizon as $x^{-1/2}$.
Such a choice of these parameters allows one to obtain a regular solution for the Green's function \eqref{Green_anzatz} throughout the entire spacetime outside the event horizon.

To determine the admissible values of the parameters $q_{0, 1}$ and $p_{0, 1}$, it is necessary to analyze the behavior of the functions $v(x)$ and $u(x)$
near the event horizon at $x = 0$. 
 To this end, we seek the solution for these functions near the event horizon in the following form:
\begin{align}
	v (x) & = v_0 + v_1 x + \cdots , 
\label{v_Taylor_1}\\
	u (x) & = u_0 + u_1 x + \cdots . 
\label{v_Taylor_2}
\end{align}
Upon substituting these expressions into Eqs. \eqref{Dirac_dim_less_1} and \eqref{Dirac_dim_less_2} and equating the coefficients of $x^{-1/2}$ and $x^0$
to zero, we obtain the following constraints on the parameters
$q_{0,1}$ and $p_{0,1}$: 
\begin{equation}
	q_0 = \frac{1}{4} (v_0 - 4 u_0 \Omega ) , \,
	p_0 = \frac{1}{4} (u_0 + 4 v_0 \Omega ), \,
	q_1 = m u_0 + v_0 , \,
	p_1 = m v_0 - u_0 . 
\label{q01_p01}
\end{equation}
Here, it should be noted that without the source terms in the form of \eqref{source_anzatz_1} and \eqref{source_anzatz_2}, the Dirac equations \eqref{Dirac_dim_less_1} and \eqref{Dirac_dim_less_2} for the two-point Green's function $G\left(t,  \vec{r} - \vec{r}_H \right)$
will not possess regular solutions. This stems from the presence of the singular terms $x^{-1/2}$
in these equations, which arise due to the spin connection $\omega_{a b \mu}$. 

\begin{figure}[H]
\begin{center}
\includegraphics[width=1.\linewidth]{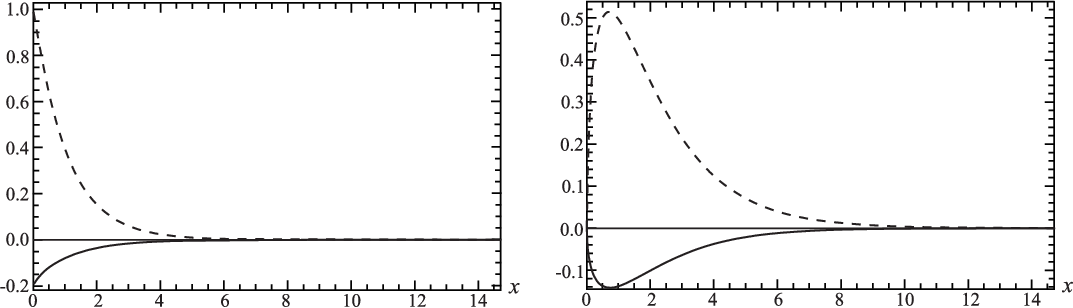}
\end{center}
\vspace{-0.5cm}
\caption{The profiles of the functions $v(x)$ and $u(x)$ for $m=1.0, \Omega = 0.409078, u_0 = 1.0, v_0 = -0.2$ (left panel)
and for $m=1.0, \Omega = 0.7197343, u_0 = 1.0, v_0 = -0.3$ (right panel). 
The solid lines represent $v(x)$, the dashed lines are for $u(x)$.
}
\label{fig_u_v_x_all}
\end{figure}

The Dirac equations \eqref{Dirac_dim_less_1} and \eqref{Dirac_dim_less_2} represent eigenvalue equations for the frequency  $\Omega$, 
which was determined using the method of successive approximations during the numerical solution of these equations. The result of the numerical computations is presented in the left panel of Fig.~\ref{fig_u_v_x_all}.

\subsection{The case of zero $v(x)$ and $u(x)$ at the event horizon
}
\label{zero}

In this case, we take {\it ad hoc} sources $j_{1,2}(x)$ in the following form: 
\begin{align}
	j_1 & = e^{-\Sigma x} \left( 
		q_1 + q_2 \sqrt{x}
	\right) , 
\label{source_anzatz_3}\\	
	j_2 & = e^{-\Sigma x} \left( 
		p_1 + p_2 \sqrt{x}
	\right) , 
\label{source_anzatz_4}
\end{align}
where $q_{1,2}$ and $p_{1,2}$ are
some arbitrary parameters chosen in such a way as to obtain a regular solution for the Green's function \eqref{Green_anzatz} throughout the entire spacetime outside the event horizon. 

As in the previous subsection, to find the admissible values of the parameters $q_{1, 2}$ and $p_{1, 2}$, it is necessary to analyze the behavior of the functions $v(x)$ and $u(x)$ 
near the event horizon. To this end, we seek the solution to the Dirac equations at $x = 0$ in the form
\begin{align}
	v (x) & = \sqrt{x} \left( v_0 + v_1 x + \cdots \right) , 
\label{v_Taylor_3}\\
	u (x) & = \sqrt{x} \left( u_0 + u_1 x + \cdots \right) . 
\label{v_Taylor_4}
\end{align}
Such a choice of the ansatz for the fermionic field allows one to eliminate the $x^{-1/2}$-type singularities on the left-hand side of the Dirac equations \eqref{Dirac_dim_less_1} and \eqref{Dirac_dim_less_2}.
Upon substituting these expressions into Eqs. \eqref{Dirac_dim_less_1} and \eqref{Dirac_dim_less_2} and equating the coefficients of $x^{-1/2}$ and $x^0$
to zero, we obtain the following constraints on the parameters
$q_{1,2}$ and $p_{1,2}$: 
\begin{equation}
	q_1 = \frac{1}{4} (3 v_0 - 4 u_0 \Omega ) , \,
	p_1 = \frac{1}{4} (3 u_0 + 4 v_0 \Omega ), \,
	q_2 = m u_0 + v_0 , \,
	p_2 = m v_0 - u_0. 
\label{q12_p12}
\end{equation}
The results of numerical calculations are presented in the right panel of Fig.~\ref{fig_u_v_x_all}. 

\subsection{Discussion of the results}

Thus, the results obtained in Secs.~\eqref{non_zero} and \eqref{zero} indicate that the Dirac equations \eqref{Dirac_dim_less_1} and \eqref{Dirac_dim_less_2} with {\it ad hoc} sources of the type \eqref{source_anzatz_1} and \eqref{source_anzatz_2} or \eqref{source_anzatz_3} and \eqref{source_anzatz_4}~-- which mimic the modified anticommutation relations for the fermionic operators $\hat{\psi}$
in curved spacetime near the event horizon~-- lead to a regular two-point Green's function $G\left(t,  \vec{r} - \vec{r}_H \right)$
describing a stationary configuration of the quantum spinor field near the black hole event horizon.

It should also be noted that the obtained solution to the Dirac equations \eqref{Dirac_dim_less_1} and \eqref{Dirac_dim_less_2} represents a Green's function
$
	G \left(t,  \vec{r} - \vec{r}_H \right) = 
	\left\langle Q \left| 
		\hat{\psi}^\dagger \left( t, \vec{r} \right) \hat{\psi} \left( t, \vec{r_H} \right) 
	\right| Q \right\rangle 
$, 
however, the question regarding the profile of the expectation value $\left\langle Q \left| \hat{\psi}(x) \right| Q \right\rangle $
remains open. The most intriguing scenario arises if $\left\langle Q \left| \hat{\psi}(x) \right| Q \right\rangle = 0$. 
 In this case, we obtain a nonperturbative fermion vacuum in curved spacetime near the event horizon, whereas the two-point Green's function $G\left(t,  \vec{r} - \vec{r}_H \right)$
describes the correlation of quantum vacuum fluctuations of the fermionic field at points $\vec{r}$ and $\vec{r}_H$; 
thus, we are dealing with the condensation of virtual ``sea'' fermions at the event horizon. Conversely, if $\left\langle Q \left| \hat{\psi}(x) \right| Q \right\rangle \neq 0$, 
 the obtained solution describes the condensation of real fermions at the event horizon.

\section{Discussion and conclusions}

Thus, the primary objective of our study was to address the following question: do other quantum effects for the behavior of fermions near the event horizon exist, apart from the Hawking radiation? We assumed that the canonical anticommutation relations for fermions must be modified in curved spacetime near the event horizon. Such a modification should alter the right-hand side (the source term) of the Dirac equation for the two-point Green's function $G\left( t, \vec{r}_H; t, \vec{r} \right) $
near the event horizon. We modeled this modification using certain functions, the explicit form of which was chosen to yield a regular Green's function near the event horizon. By solving the Dirac equation with a source for the two-point Green's function as an eigenvalue problem for the spinor frequency $\Omega$,
 we demonstrated that a regular solution exists throughout the entire spacetime outside the event horizon.

Another argument in favor of the coexistence of an event horizon and a regular fermionic field is that both black holes and fermions exist in nature; therefore, fermionic fields must be regular in the vicinity of a black hole event horizon. In particular, an expression for the two-point Green's function of the fermionic field must exist.


\begin{thebibliography}{99}

\bibitem{Hawking:1974rv}
S.~W.~Hawking,
Nature \textbf{248}, 30 (1974).

\bibitem{Maggiore:1993rv}
M.~Maggiore,
Phys. Lett. B \textbf{304}, 65 (1993). 


\bibitem{Maggiore:1993kv}
M.~Maggiore,
Phys. Lett. B \textbf{319}, 83 (1993) . 


\bibitem{Capozziello:1999wx}
S.~Capozziello, G.~Lambiase, and G.~Scarpetta,
Int. J. Theor. Phys. \textbf{39}, 15 (2000). 


\bibitem{Chandrasekhar}
S. Chandrasekhar,
{\it The mathematical theory of black holes} (Oxford University Press, New York, 1998).

\bibitem{Malik:2024nhy}
Z.~Malik,
Annals Phys. \textbf{479}, 170046 (2025). 


\bibitem{Finster:1998ws}
  F.~Finster, J.~Smoller, and S.~T.~Yau,
  Phys.\ Rev.\ D {\bf 59}, 104020 (1999).  
  
  \bibitem{Herdeiro:2017fhv}
    C.~A.~R.~Herdeiro, A.~M.~Pombo, and E.~Radu,
    Phys.\ Lett.\ B {\bf 773}, 654 (2017).

\bibitem{Herdeiro:2019mbz}
C.~Herdeiro, I.~Perapechka, E.~Radu, and Y.~Shnir,
Phys. Lett. B \textbf{797}, 134845 (2019).
    
 \bibitem{Konoplya:2021hsm}
 R.~A.~Konoplya and A.~Zhidenko,
 Phys. Rev. Lett. \textbf{128},  091104 (2022).

\bibitem{Dzhunushaliev:2025lki}
V.~Dzhunushaliev, V.~Folomeev, N.~Beissen, and A.~Nurmukhamedov,
Eur. Phys. J. C \textbf{86},  240 (2026).

\bibitem{Dzhunushaliev:2025ntr}
V.~Dzhunushaliev and V.~Folomeev,
Gen. Rel. Grav. \textbf{58},  39 (2026).

\bibitem{Dzhunushaliev:2025dma}
V.~Dzhunushaliev and V.~Folomeev,
Phys. Lett. B \textbf{876}, 140416 (2026).

\bibitem{heis}
W. Heisenberg, \textit{Introduction to the unified field theory of elementary particles} (Max-Planck-Institut f\"ur Physik und Astrophysik, Interscience Publishers London, New York, Sydney, 1966).
  
\bibitem{Lawrie2002}
I.~Lawrie, {\it A Unified Grand Tour of Theoretical Physics} (Institute of Physics Publishing, Bristol, 2002).


\end{thebibliography}
\end{document}